\documentclass[preprint2]{aastex}
\input{psfig}

\newcommand{\nh}{N_{\rm H}}

\newcommand{\pica}{{\it Pictor A~}}

\newcommand{\xmm}{{\it XMM-Newton~}}
\newcommand{\chandra}{{\it Chandra~}}

\def \nh {N${\rm _H}$~}

\begin{document}

\title{Detection of X-ray Emission from the Eastern Radio Lobe of  PICTOR A}

\author{Paola Grandi}
\affil{Istituto di Astrofisica Spaziale e Fisica Cosmica -- CNR,
Sede di Bologna, Via Gobetti 101, I-40129 Bologna, 
Italy}
\email{grandi@bo.iasf.cnr.it}
\author{Matteo Guainazzi} 
\affil{\xmm Science Operation Center, Madrid, Spain}
%\email{mguainaz@xvsoc01.vilspa.esa.es}
\author{Laura Maraschi} 
\affil{ Osservatorio Astronomico di Brera, Milano, Italy}
%\email{maraschi@brera.mi.astro.it}
\author{Raffaella Morganti}
\affil{Netherlands Foundation for Research in Astronomy, Dwingeloo, 
The Netherlands }
%\email{morganti@astro.nl}
\author{Roberto Fusco-Femiano}
\affil{Istituto di Astrofisica Spaziale e Fisica Cosmica -- CNR,
Roma, Italy}
%\email{dario@ias.rm.cnr.it}
\author{Mariateresa Fiocchi}
\affil{Asi Scientific Data Center, Frascati, Italy}
%\email{mariateresa.fiocchi@est.asi.it}
\author{Lucia Ballo and Fabrizio Tavecchio} 
%\affil{ Osservatorio Astronomico di Brera, Milano, Italy}
%\email{luballo@brera.mi.astro.it}
%\author{Fabrizio Tavecchio} 
\affil{ Osservatorio Astronomico di Brera, Milano, Italy}
%\email{fabrizio@brera.mi.astro.it }

\begin{abstract}
The \xmm satellite has revealed extended X-ray emission from the 
eastern radio lobe of the Fanaroff-Riley II Radio Galaxy \pica.
The X-ray spectrum, accumulated on a region covering about half the
entire radio lobe, is well described by both a thermal ($kT=5^{+9}_{-2}$ keV) model
and  a power law with an energy index $\alpha_{X-ray}=0.6\pm0.2$.
The X-ray emission could be thermal and produced by circum-galactic gas 
shocked by the expanding radio lobe or, alternatively, by Inverse Compton 
(IC) of cosmic microwave background photons by relativistic electrons in 
the lobe. The latter possibility seems to be supported by  
the good agreement between the  lobe-average synchrotron 
radio index ($<\alpha_{radio}>$=0.8) and the X-ray energy slope 
$\alpha_{X-ray}$.
% and by the good
%spatial coincidence of the radio and X-ray emission.
However, if this is the case, the magnetic field ($B_{IC}\sim1-2\mu$ G), 
as deduced from the comparison of the IC X-ray and radio fluxes, 
is more than a factor 2 below the equipartition  
value
 %($B_{eq}\sim 5 \mu$ G) 
estimated in the same X-ray region.

\end{abstract}


\section{Introduction}
The recent opportunity offered by \chandra and \xmm 
to perform  spatially resolved studies  of radio loud AGN have focused 
attention on X-ray processes occurring at pc-kpc distance from the nucleus 
(Kraft et al. 2000, Sambruna et al. 2000, Wilson et al. 2000, Tavecchio 
et al. 2001, Harris \& Krawczynski, 2002, Hardcastle et al. 2002).
This is an interesting new perspective, as  the past X-ray studies were
almost completely dedicated to the investigation of the nuclear emission.

In the environment of radio loud AGN there are several
structures which emit X-ray photons: jets, hot gaseous media, 
hot spots and extended radio lobes. Among these, radio lobe counterparts,
are the less bright and therefore the most difficult regions to be detected. 
This is a frustrating condition, as the X-ray detection of radio extended structures could provide, in 
principle,  important physical information.
It is thought that radio-emitting electrons 
can upscatter  local cosmic microwave background (CMB) 
and/or nuclear AGN photons, producing X-rays 
(Feigelson et al. 1995, 
Kaneda et al. 1995,  Brunetti et al. 1999, 2001a, 2001b, 
Hardcastle et al 2002).
Measures of Inverse Compton (IC) 
radiation provide a formidable instrument to probe the energetics 
of the radio lobes, i.e. the integrated energy channelled by the 
active nucleus in the jet during its lifetime. 
In particular, if observed, X-ray photons allow a direct estimate 
of the average magnetic field along the line of sight
and the energy densities of the particles in the lobe.
However this is not an easy task.
The corresponding X-ray fluxes are expected to be weak and 
easily overwhelmed by thermal emission surrounding many radio galaxies.
In addition, external gas compressed by the radio lobes during their 
expansion through the environment could also contribute 
to the X-ray emission.
Disentangling thermal and non-thermal radiation is then more complicated.

In spite of the difficulty, some 
X-ray detections of extended synchrotron radio regions have 
been already realized (Feigelson et al. 1995, Kaneda et al. 1995, 
Tahiro et al. 1998, 2001, Brunetti et al. 1999, Brunetti et al. 2001a, 
Brunetti et al. 2001b, Hardcastle et al. 2002) and for the most part the 
more plausible mechanism for the X-ray production seems to be Inverse Compton.

Here we present a \xmm detection of X-ray photons from 
the east radio lobe of \pica. 
\pica is a nearby (z=0.035) radio galaxy optically classified 
as Broad Line Radio Galaxy.
It is an isolated source. As reported by Miller et al. (1999), 
who performed an accurate X-ray optical  study of the environment 
around nearby radio galaxies, 
there is no indication of galaxy clustering around this source.
\pica is a double lobed radio source with a Fanaroff Riley II (FRII) 
morphology.  VLA observations of Perley et al. (1997; PRM) show 
two nearly circular radio lobes 
with hot spots.  The western lobe has a flatter radio spectrum and 
a much brighter hot spot. The spectrum indices are however fairly uniform 
throughout each lobe $<\alpha_r=0.8>$, although a 
flattening is observed  near the hot spots. 
A faint radio jet connects the nucleus to the western hot spot.

Recently, \chandra has produced X-ray spectra of 
the jet and of the western hot spot and detected faint extended
emission from the east lobe (Wilson et al. 2001). 
\xmm, collecting a larger number of photons, has allowed us not only to detect the extended emission from the lobe, but also to obtain a spectrum 
between 0.2-12 keV.
We think that this exciting result has been possible thanks 
to the combination of four important factors: i) the very large 
effective area of \xmm, which allows to reveal X-ray structures of 
low surface brightness;  ii) the spatial resolution 
of \xmm which can resolve kpc structures in sources with moderately 
low redshifts;  iii) the high brightness of the radio lobes of \pica;
iv) the lack of a dense hot intracluster  medium, which can mystify the 
X-ray measures. 
% being \pica an isolated FRII radio galaxy.

Here we will essentially focus on the emission of the eastern lobe.
The other X-ray components and in particular the nucleus  will be treated 
in a following paper (Grandi et al. in preparation).

\section{\bf Data Analysis and Results}

The field containing \pica  was observed by \xmm (Jansen et al. 2001)
on March 17, 2001. 

The EPIC/p-n  cameras (0.1-15~keV; Str\"uder et al. 2001) operated in
Full Frame Mode, with the Thin optical  filter.
The EPIC/MOS cameras (0.25-10~keV; Turner et al. 2001) 
operated in Small Window Mode. In this mode only a 1.8'$\times$1.8' 
square of the field of view is active in order to avoid pile up 
problems with the nucleus.
The spectral analysis of the environment 
of \pica is therefore based on the p-n instrument alone.

Data reduction was performed using the software package {\sc SAS} 
version 5.1. A Calibration Index File appropriate for the date 
of the observation and the data analysis (July 2001) was produced. 
We employed the widest pattern selection ({\tt PATTERN=0-4 }: singles 
and doubles events), for which the response matrices are available.
Different choices of the pattern selection did not
substantially affect the following results.
The observation was affected by flaring background.
The inspection of the p-n light curve showed an
irregular activity, concentrated in the central part and towards
the end of the observation. 
We verified that the spectral fit results were not substantially affected by 
the  inclusion of the high-background intervals.
As the signal-to-noise ratio substantially increases if the whole p-n data 
set is used,  we did not apply any time screening to the p-n data, beyond those automatically applied by the data reduction pipeline. 
After screening, the total exposure time was 15.7~ks. 

\subsection{\it X-ray Image}
Figure 1 ({\it upper panel}) shows the 0.2-12 keV image of \pica.
Several components can be revealed by a simple inspection of
the \xmm field: the bright nucleus, the western hot spot, the jet 
pointing towards the hot spot and a faint asymmetric diffuse emission 
around the nucleus. The point-like sources at about 3 arcmin NE 
from the nucleus and  at the lower extremity of the east lobe 
are foreground objects.

\begin{figure}
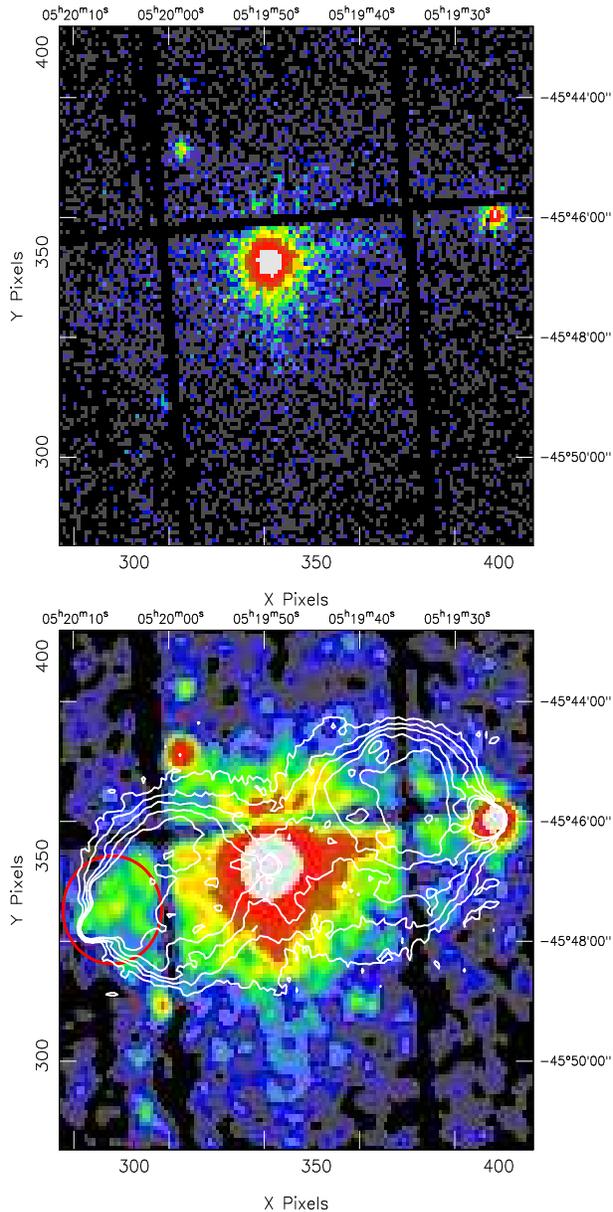

\begin{center}
\vbox{
\psfig{figure=figure1a.ps,height=8.0cm,width=8.0cm,angle=-90}
\psfig{figure=figure1b.ps,height=8.0cm,width=8.0cm,angle=-90}
}
\end{center}
\vspace{-0.1cm}
\caption{\footnotesize{{\bf Upper Panel}: XMM/p-n image of \pica in the 0.2--12~keV energy.
The field of view is 10.4 $\times$ 10.4 arcmin and the pixel 
dimension is 4 arcsec. This image was accumulated on time intervals not 
affected by high background flares. 
Several components can be revealed by a simple inspection of
the field of view: the  bright nucleus, the western hot spot, the jet 
pointing towards the hot spot and a faint nebulosity extending around and 
eastwards from the nucleus. The east extended structure coincides with a 
synchrotron radio  lobe. 
The two  point-like sources in the field at $\alpha(2000)=05^{h} 19^{m} 58.5^{s}$, 
$\delta(2000)=-45^{\circ} 44' 52.0''$  
and $\alpha(2000)=05^{h} 20^{m} 00.1^{s}$, 
$\delta(2000)=-45^{\circ} 49' 00.1''$ are foreground objects. {\bf Lower panel}: Radio contours from a 20 cm (1.4 GHz) radio VLA map with 
$7.5$ arcsec of resolution (PRM) are plotted on 
the same p-n image smoothed with a Gaussian 
with $\sigma=1.5$ pixel (6'' arcsec). 
The X-ray and radio images 
have been re-aligned because of a relative shift. 
The east extended X-ray structure coincides  with the synchrotron radio lobe.
The red circle shows the region were the X-ray analysis has been 
performed.}
}
\label{fig1}
\end{figure}

In Figure 1 ({\it lower panel}),  the 20 cm (1.4 GHz) radio contours is 
overimposed on the  \xmm image smoothed with a gaussian with $\sigma=6''$.
The faint nebulosity becomes more evident showing an 
asymmetric and elongated shape. The spatial coincidence of the 
eastward structure and the radio lobe is impressive.
On the contrary, the X-ray counterpart of the west lobe is less clear.
The presence of a strong hot spot and a jet emission probably reduces 
the contrast between the faint extended emission and the background.
In addition, we have lost most of the western X-ray lobe 
because of the gaps among the CCD cameras.

We exclude the possibility that unresolved point-sources produce the eastwards
diffuse X-ray emission.
The high spatial resolution of \chandra has  revealed four 
faint compact objects in the eastern lobe.  We retrieved the 
\chandra/ACIS public data  and estimated an average flux 
of F$_{\it 0.2-2~keV}\sim2-3\times 10^{-15}$  erg cm$^{-2}$ sec$^{-1}$ 
for each source, assuming a power law model ($\Gamma=1.7$) and 
Galactic absorption. 
Then, about 20 point-like sources would be necessary to reproduce 
the (Galactic absorbed) X-ray enhancement observed by \xmm  
(F$_{\it 0.2-2~keV}=5.9\times 10^{-14}$  erg cm$^{-2}$ sec$^{-1}$).

\subsection{\it  X-ray Spectrum}

We accumulated a X-ray spectrum of the east lobe 
on a circular region of radius 54'' (26 kpc\footnote{1'' corresponds to 
483 pc for H$_0 = 100$~km~sec$^{-1}$~Mpc$^{-1}$} 
located at a distance of 172'' (83 kpc) from the 
nucleus (red circle in Figure 1-{\it lower panel}). 
This corresponds to the lobe area characterized by intense X-ray and radio
emission. 
The background was extracted from a circle of the same 
dimension  in the same chip. 
Before performing the fits, the spectrum was binned in order to ensure
the applicability of the $\chi^2$ statistic and an adequate sampling 
of the spectral resolution of each instrument.
All the fits were performed with XSPEC11.

In the 0.2-12 keV band, the net count rate in the 
X-ray lobe area, after the removal 
of the background, was $5.6\times10^{-2}$ with a 1-$\sigma$ 
uncertainty of $0.5\times10^{-2}$.  The signal to noise ratio 
larger than 11 excludes the possibility that the 
measured X-ray excess was a background fluctuation.

\begin{figure}
%\begin{center}
\vbox{
\psfig{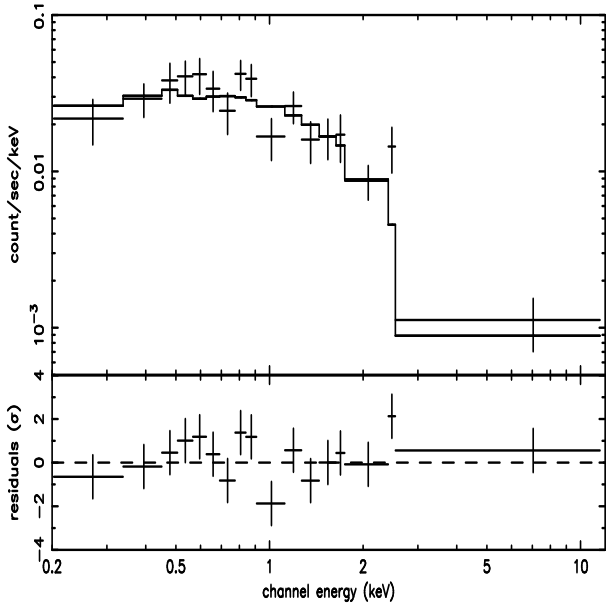}}
%\end{center}
\vspace{-0.2cm}
\caption{\footnotesize{
\xmm pn spectrum when a power law 
is fitted to the data. The column density is fixed at the Galactic value
\nh=$4.2\times10^{20}$ cm$^{-2}$ }}
\end{figure}
 
The X-ray spectrum was well reproduced either by a power law or by 
a thermal model in the 0.2-12 keV.
In both cases, the acceptable range of \nh  was consistent with the 
Galactic line of sight value \nh=$4.2\times10^{20}$ cm$^{-2}$.
Then, we fixed the column density to the Galactic value in order to reduce 
the uncertainties of the parameters. The results are listed in Table 1.
The reported uncertainties are at 90$\%$ confidence level for 
one parameter of interest ($\Delta\chi^2=2.71$).

It is worth noting that the spectral energy index ($\alpha_x = \Gamma -1$)
of the power law is in very good agreement with the average 
synchrotron radio index of the lobes ($< \alpha_r > = 0.8$).
This is an intriguing result that suggests a common non-thermal origin of
radio and X-ray photons.
However, as discussed later, this is not the only possibility.

\begin{deluxetable}{lcclc}
\tabletypesize{\small}
\tablewidth{0pc}
\tablecolumns{4}
\tablecaption{\xmm fits to the Eastern Lobe of \pica in the 
0.2-12 keV range. \nh is fixed to the Galactic value = 4.2$\times10^{20}$ 
cm$^{-2}$}
%\tablehead{\colhead{} &\colhead{}&\colhead{}&\colhead{}}
\startdata
\multicolumn{2}{l}{Power Law}&&\multicolumn{2}{l}{Thermal Emission}\\
&&&\\
$\Gamma$ & 1.56$^{+0.20}_{-0.18}$ &&kT [keV]  & $5^{+9}_{-2}$\\ 
N$^a$    & 2.5$^{+0.4}_{-0.2}$    &&$I_{\rm s}^b$ &  $1.6_{-0.2}^{+0.2}$\\
         &                        &&A$^c$ & $0<0.7$\\  
$\chi^2$(dof)  & 59(58)            &&$\chi^2$(dof)    &60(57)\\
Flux$^d$ (0.2-2 keV)& $7.8^{+1.0}_{-0.5}$ &&Flux$^d$ (0.2-2 keV)
&$7.6^{+1.0}_{-1.0}$\\
Flux$^d$ (2-10  keV)& $12.7^{+1.0}_{-2.0}$  &&Flux$^d$ (2-10  keV)&
$8.8^{+1.0}_{-1.0}$ \\
\enddata
\tablenotetext{a}{Normalization at 1 keV ($\times10^{-5}$ photons 
cm$^{-2}$ s$^{-1}$ $keV^{-1}$). }
\tablenotetext{b}{Emission measure in $10^{64}$ cm$^{-3}$.}
\tablenotetext{c}{Metal abundances with the same ratio to the solar abundance .}
\tablenotetext{d}{Unabsorbed flux in units of 10$^{-14}$ erg  cm$^{-2}$ s$^{-1}$}
\end{deluxetable}

\section{Discussion}

In this paper we report a \xmm spectral analysis of an extended 
radio structure. This is the first time that \xmm detects and produces a 
spectrum of a radio lobe in an isolated powerful FRII radio galaxy. 
The fit result is not univocal: both a non-thermal 
and  a thermal model are acceptable from a statistical point of view.
In the following we investigate both these possibilities.

\noindent
{\it Thermal emission} - Although \chandra (Wilson et al. 2001) and 
\xmm (paper in preparation) 
have revealed a gaseous atmosphere of about 1' of radius around the nucleus,
we exclude that the X-ray lobe emission is part of this circumgalactic medium.
The studied region is located far away from the nucleus, at about 2' and, 
in addition, its temperature is much harder than the gas surrounding 
the nucleus (kT$\sim 0.3$ keV; paper in preparation).

It is also improbable that the lobe is completely 
filled by gas  gathered from the nuclear environment. 
Recent  \chandra images of clusters show very low density regions
in correspondence to the radio lobes of 
the central galaxies (Fabian et al. 2001, 
Finoguenov and Jones 2001, McNamara et al. 2000). 
It is evident  that radio lobes escaves cavities during their expansion in the
surrounding medium.
Moreover, the  Rotation Measure 
(RM= 43\footnote{RM=$\propto n_e B_\parallel L$, 
where $n_e$ is the electronic density (cm$^{-3}$), $B\parallel$ (gauss)
is the Magnetic Field along the line of sight and L is the path length 
(in parsec)}) reported by PRM seems 
to be inconsistent with  the idea that synchrotron and thermal 
plasma (causing Faraday rotation) coexist within the east lobe.
As deduced by the emission measure in Table 1, the gas, 
if uniformly distributed in the X-ray region, should have a density of 
$n_e^{\it X-ray}\sim 3\times 10^{-3}$ cm$^{-3}$.
Assuming an equipartition magnetic field of B$\sim 5\mu$G (see later), 
the observed measure of rotation requires a thermal plasma density 
smaller by at least a factor 10.

Nevertheless the thermal hypothesis can not be rejected.
The X-ray radiation could be actually  produced by shocked gas 
surrounding the radio lobe. (This could also explain the
strong radio depolarization observed in the eastern lobe at 90 cm (PRM)).
Although 
\chandra images show cool rims rather than strong 
shocks around the radio lobes in clusters (Fabian et al. 2000, McNamara, 
Blanton et al. 2001),  theoretical arguments support this view.
Hydrodynamic simulations show that, in the early life of 
a powerful source,  jet materials inflate overpressured cocoons 
which compress the surrounding 
ambient material. At a later time, the shocks become weaker and 
eventually, when the jet activity ceases, the 
old cocoon material can rise under the action of 
buoyancy forces, gather cold gas from the
from the surrounding ambient and lift them up into the galaxy/cluster 
atmosphere (Reynolds et al. 2002).
Then,  the observed cool rims surrounding the radio cavities 
in clusters could then be remnants of old cocoons.
On the contrary, \pica  could be  in an early stage.
Its  cocoons are still supersonic and able to drive  a bow shock into the 
circumgalactic gas.
Zanni et al. (2002) have recently studied the evolution of an expanding cocoon 
as a function of the jet physical properties and shown that
high Mach number jets with high densities (with  respect to the environment) 
are strongly overpressured for a long fraction of their life.
%Their also show that, in this case, shells of enhanced X-ray 
%emission surround the radio lobes (as observed) rather than deficit of X-ray 
%photons.

\noindent
{\it Non-Thermal emission} - The good agreement between the  
lobe-average synchrotron 
radio index ($<\alpha_{radio}>$=0.8, PRM) and the power law slope
($\alpha_{X-ray}$) surely favours the idea that 
radio and X-ray photons are produced by the same relativistic 
population of electrons.
At first, we checked whether the synchrotron process, 
responsible for the radio emission, could also produce X-rays. 
In order to verify that, we produced a radio-X-ray spectrum of the east lobe.
We directly estimated the flux of the X-ray region 
from the 20 cm radio map (PRM). The contribution of the (double) hot spot 
was removed and the resultant radio flux ($\sim 3.2$ Jy ) compared to that 
from the entire lobe. Successively, the radio fluxes at 74, 327 and 4995 MHz,  
listed in Table 3 of RMP,  were rescaled by the same factor ($\sim 12\%$).
As shown in the upper panel of Figure 3, 
our X-ray data do not lie on the extrapolation of 
the radio spectrum. Moreover they can not be connected to the radio invoking
radiative losses of the relativistic electrons, because
the \xmm spectral slope is hard and consistent with the
radio one.  Synchrotron process is implausible.

\begin{figure}
%\begin{center}
\vbox{
\psfig{figure=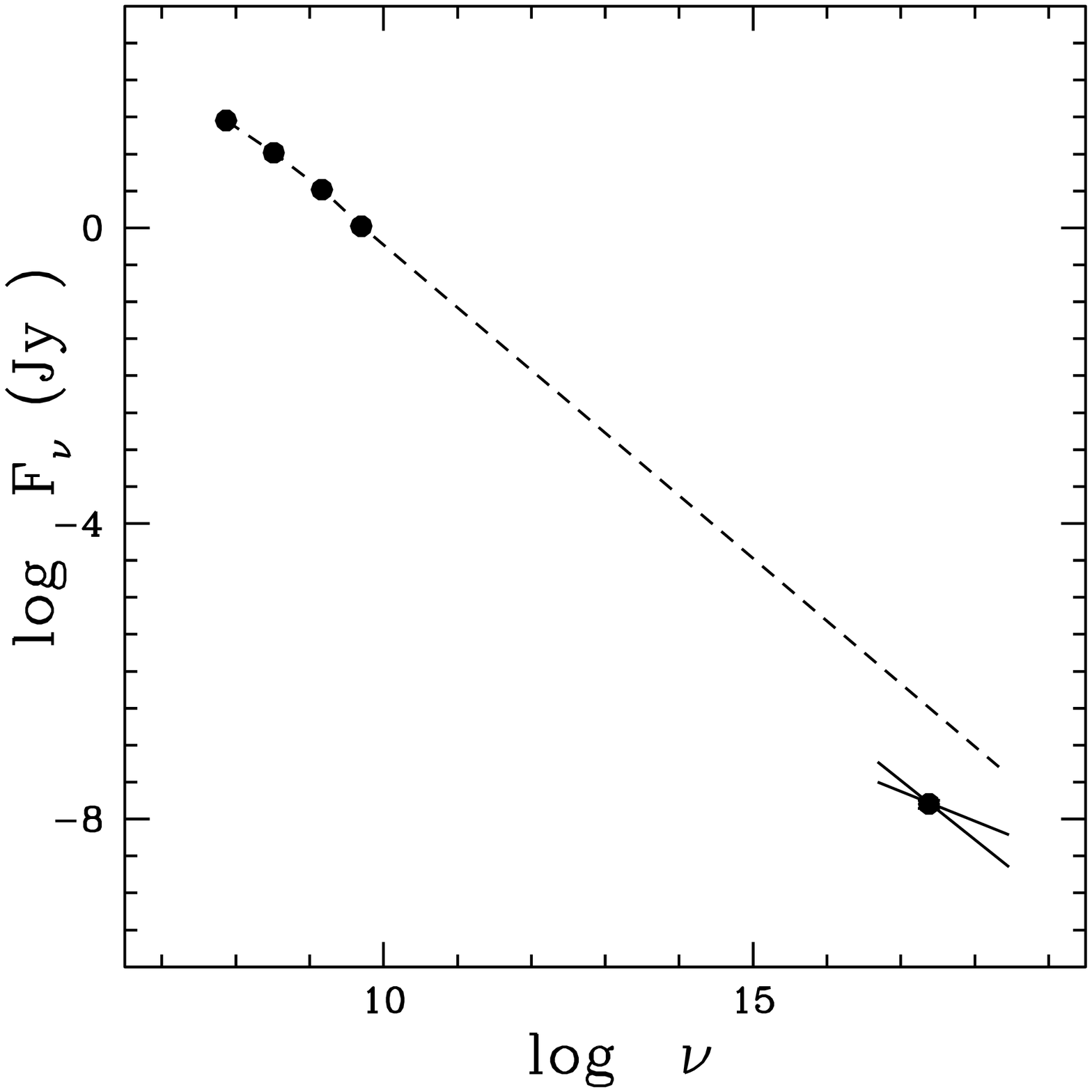,height=8.0cm,width=8.0cm}
\psfig{figure=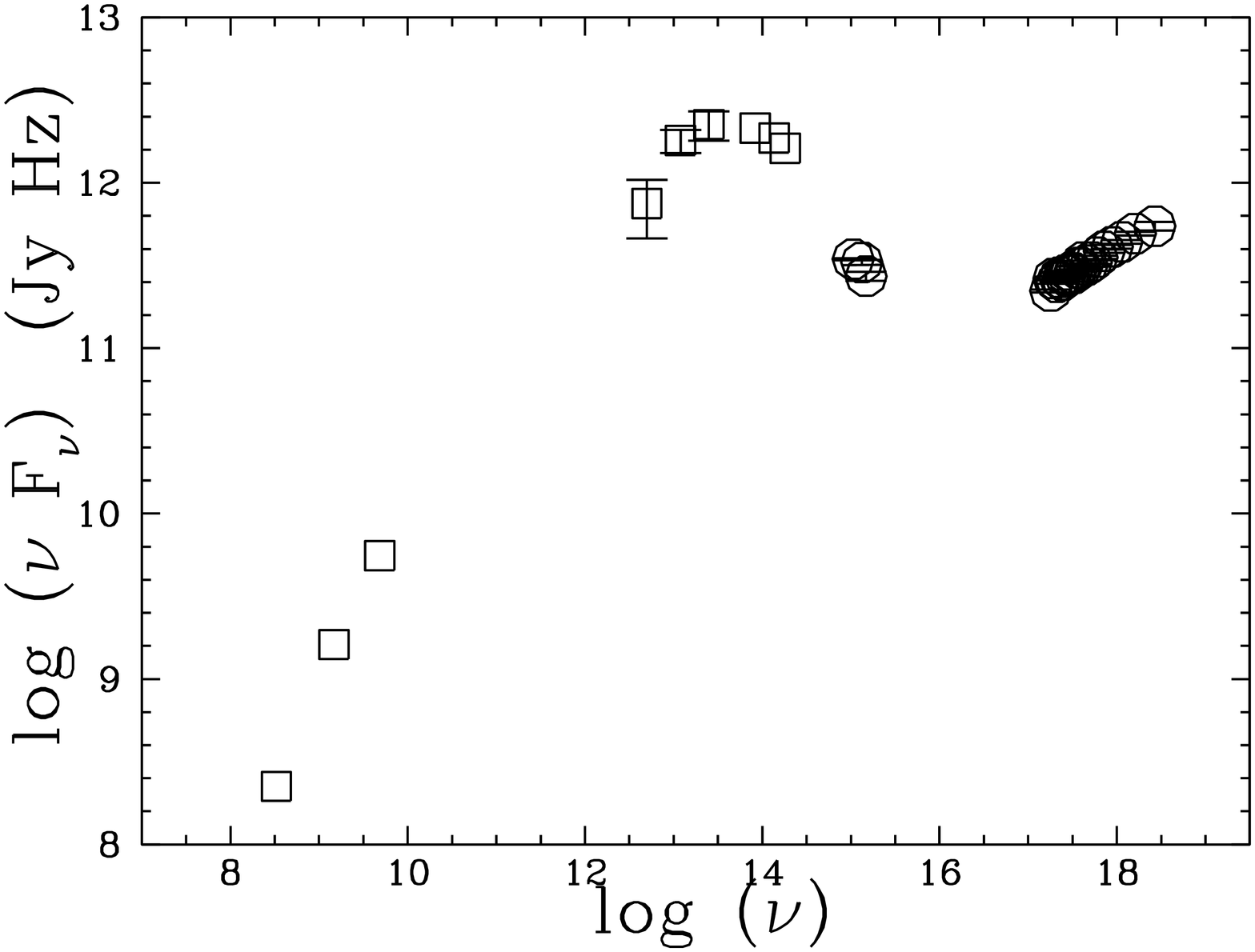,height=8.0cm,width=8.0cm,angle=-90}}
%\end{center}
\caption{\footnotesize{{\bf Upper Panel} 
The radio X-ray spectrum of the east lobe of \pica.
The radio data are from PRM and 
the \xmm spectrum is from this paper. The radio fluxes have been 
normalized to the region covered by the X-ray emission (see text).
{\bf Lower Panel} Spectral energy distribution of the nucleus of \pica, 
from non-simultaneous observations. Radio-infrared data
(open squares) are taken from literature. Simultaneous UV and X-ray 
data (open circles) refer to the XMM observation on 2001 March 17.}}
\end{figure}

Thus, the Inverse Compton remains the more likely emission process. 
The seed photons scattered off by the lobe electrons can have 
different origins.
They can be the same synchrotron photons produced in the lobe (SSC model), 
photons coming from the AGN in the \pica nucleus (IC/AGN) and cosmic microwave
background photons (IC/CMB).
In order to discriminate among them, it is necessary to 
estimate the relative energy densities.
Extrapolating the radio spectrum in Figure 3 (upper panel) from $10^6$ Hz 
to 10$^{12}$ Hz, we inferred 
a synchrotron photon density of $u_{\it lobe}\sim 10^{-15}$ 
erg cm$^{-3}$ (in the X-ray region).
%Even extending the synchrotron flux integration up to $10^{15}$
%Hz (corresponding to an improbable synchrotron spectrum without
%radiative losses up to the UV band), the energy density is $\sim 6\times10^{-1%5}$ erg cm$^{-3}$. 
In comparison, the CMB energy density at the \pica redshift is
$u_{CMB}=4.7 \times10^{-13}$ erg cm$^{-3}$, 
about 2 order of magnitude larger.

Analogously we estimated the AGN  energy density 
at a distance of 83 kpc, where the extracted X-ray region 
is located, utilizing the spectral energy distribution (SED) of 
the nucleus of \pica.
The nuclear SED, shown in Figure 3 (lower panel), is the combination of
the UV fluxes and the X-ray fluxes measured with this XMM observation 
(Grandi et al. in preparation) and radio and infrared data from literature 
(PRM, Golombek et al. 1988, Glass 1981).

Even assuming the implausible case in which all the radio-UV nuclear 
photons are IC seed photons, the energy density is 
$u_{\it AGN}\sim  5 \times 10^{-15}$ erg cm$^{-3}$, again 
significantly smaller than the CMB radiation. 

Thus, the Inverse Compton of the cosmic microwave background 
photons by the relativistic electrons in the east lobe is 
the more probable non-thermal mechanism responsible for 
the observed X-ray flux.

As an immediate consequence, we can directly estimate 
the magnetic field ($B_{\it IC}$) in the analyzed lobe region, 
following the prescription of  Harris $\&$ Grindley (1979).
Assuming F$_{1.4~GHz} =3.2$ Jy, 
F$_{0.2-2~keV} =7.8\times10^{-14}$ erg cm$^{-2}$ sec$^{-1}$ 
and $\alpha=0.8$, we obtain a magnetic field of  
$B_{\it IC}\sim 2\mu$ G. 
This value is smaller by about a factor 2.5 than 
the equipartition field (B$_{eq}\sim 5 \mu$ G) inferred by the standard 
formula reported by Miley (1980).
Our input parameters were: F$_{1.4~GHz} =3.2$ Jy and $\alpha=0.8$ (as before),
$\theta=108''$, s=52.18 kpc, k=1, $\eta$=1, $\phi=90$, 
where $\theta$ is the X-ray diameter region, s is the path length, 
k is the ratio of the energy in heavy particles to the energy 
in electrons, $\eta$ is the volume filling factor and $\phi$ is the angle 
between the uniform magnetic field and the line of sight.
(Note that we obtained a similar value of B$_{eq}$  considering the
entire lobe and the total radio flux).
If we consider a harder slope of the electron energy 
distribution ($N \propto \gamma^{-p}$) in agreement with 
the X-ray energy index ($\alpha_x=0.6$, p$=2\alpha+1$), the estimated 
magnetic field is only $B_{\it IC} \sim 1 \mu$ G and the violation of the
equipartition condition becomes more serious.
%An IC/CMB model able to reproduce 
%the \xmm flux (see Figure 3) requires then an electron energy distribution 
%of $u_e\sim 4 \times 10^{-11}$ erg cm$^{-3}$ 
%($\gamma_{min}=10^3$), significately larger 
%than the magnetic density $u_B=4\times 10^{-14}$ erg cm$^{-3}$.  
%Note that the difference between the
%electron and magnetic densities further increases 
%if $\gamma_{min}$ is shifted to lower energies.  

Then,  if the X-ray emission is due to scattering of the microwave background 
radiation, the radio lobe could be far from the equipartition condition.
This results is not surprising. Recently Hardcastle et al. (2002) found
that the X-ray emission from radio lobes of three radio sources 
(two quasars and
one radio galaxy), if due to IC/CMB, needed IC magnetic fields 
typically a factor 2 below the 
equipartition values.

However, we would like to stress that the equipartition and IC methods 
used to determinate the magnetic field are not accurate.
In addition, our radio and X-ray measurements are affected by 
uncertainties. One of these is the X-ray contribute of the radio hot spots, 
that we can not quantize (if present) with the current XMM-Newton image.

\section{Conclusion}

\xmm has revealed extended X-ray emission coincident with the east 
radio lobe of \pica, an isolated powerful FRII radio galaxy.
%Our data can not allow us to reject neither a thermal nor a non-thermal 
%origin of the X-ray photons.

The X-ray radiation could be produced by material of the circum-galactic halo 
which is swept and heated by a bow shock in front of the radio lobe.
%In favour of this interpretation there is a strong depolarization of the 
%eastern lobe observed at 90 cm by Perley et al. (1997).
In disfavour of this interpretation there are 
some recent high resolution images of Chandra
which show no strong shocks around the radio lobes (i.e. the empty cavities) 
in clusters. 
However, as supported by hydrodynamic simulations, 
\pica  could be in an early phase of its evolution characterized by 
still overpressured cocoons. 
On the contrary, the radio galaxies in clusters could be at a 
final stage with the old (decelerated) cocoons under the only
action of buoyancy forces.

Alternatively, the X-ray emission could due to a non-thermal mechanism.
The good agreement between the X-ray energy spectral index 
($\alpha_X=0.6\pm0.2$) and the average value of the synchrotron radio 
slope seems to support  this possibility.  
The more likely non-thermal process is Inverse Compton scattering of 
the local CMB photons by relativistic electrons in the lobe.
However, if this is the case, the IC magnetic field, estimated in the X-ray region, 
turns out to be a factor 2.5 or more below the equipartition  value.
%If the $u_e$ and $u_B$ energy densities are forced to be similar, 
%the IC/CMB process is not efficient enough to produce all the observed X-ray 
%photons. 

\acknowledgements
We are very grateful to Dr. Perley for kindly providing the radio map.
We also wish to thank D. Harris for very stimulating 
e-mail comments and suggestions and L. Feretti, G. Brunetti, M. Murgia and C.
Stanghellini for useful discussions.
We are particularly grateful to the referee, K. Meisenheimer, for very  
constructive suggestions.

\end{document}